\documentclass{article}
\usepackage{arxiv}

\usepackage[utf8]{inputenc} % allow utf-8 input
\usepackage[T1]{fontenc}    % use 8-bit T1 fonts
\usepackage{url}            % simple URL typesetting
\usepackage{booktabs}       % professional-quality tables
\usepackage{amsfonts}       % blackboard math symbols
\usepackage{nicefrac}       % compact symbols for 1/2, etc.
\usepackage{microtype}      % microtypography
\usepackage{graphicx}
\usepackage{afterpage}
\usepackage{placeins}
\usepackage{cite}
\usepackage{amsmath,amssymb,amsfonts}
\usepackage{algorithmic}
\usepackage{graphicx}
\usepackage{textcomp}
\usepackage{mathtools}
\usepackage{subcaption}
\usepackage{multirow}
\usepackage{lipsum}
\usepackage{titlesec}
\usepackage{amsmath}
\usepackage[table,xcdraw]{xcolor}
\usepackage{array}
\usepackage{longtable}
\usepackage{balance}
\usepackage{hyperref}
\usepackage{doi}

\title{RSSI Estimation for Constrained Indoor Wireless Networks using ANN}

\author{Samrah Arif*,
        M. Arif Khan*,
        and Sabih ur Rehman*
        \\ % <-this % stops a space\\
        {*School of Computing, Mathematics and Engineering, Charles Sturt University, Australia}}
\begin{document}
\maketitle
\begin{abstract}
	In the expanding field of the Internet of Things (IoT), wireless channel estimation is a significant challenge. This is specifically true for low-power IoT (LP-IoT) communication, where efficiency and accuracy are extremely important. This research establishes two distinct LP-IoT wireless channel estimation models using Artificial Neural Networks (ANN): a Feature-based ANN model and a Sequence-based ANN model. Both models have been constructed to enhance LP-IoT communication by lowering the estimation error in the LP-IoT wireless channel. The Feature-based model aims to capture complex patterns of measured Received Signal Strength Indicator (RSSI) data using environmental characteristics. The Sequence-based approach utilises predetermined categorisation techniques to estimate the RSSI sequence of specifically selected environment characteristics. The findings demonstrate that our suggested approaches attain remarkable precision in channel estimation, with an improvement in MSE of $88.29\%$ of the Feature-based model and $97.46\%$ of the Sequence-based model over existing research. Additionally, the comparative analysis of these techniques with traditional and other Deep Learning (DL)-based techniques also highlights the superior performance of our developed models and their potential in real-world IoT applications.
\end{abstract}

% keywords can be removed
\keywords{Wireless Channel Estimation \and Low-Power IoT \and Deep Learning \and Neural Networks}

\section{Introduction}
The Internet of Things (IoT) has introduced a new phase of interconnectedness, allowing various objects to communicate and interact with one another through wireless networks \cite{Ammar2019}. In such interconnected environments, accurate channel estimation is critical, especially in the context of LP-IoT applications \cite{Kang2023}. Wireless channel estimation is essential for ensuring resilient communication, particularly in LP-IoT applications where limited resources require both efficiency and accuracy. Traditional approaches for channel estimation, such as Least Squares (LS) and Maximum Likelihood (MaxL), often need to catch up in accurately capturing the intricate features of wireless channels in real-world situations, especially under diverse conditions \cite{Wu2020}. This gap prompts a shift towards more adaptable and intelligent wireless channel estimation solutions for LP-IoT networks. In this context, Machine Learning (ML) has increasingly been recognised as a potential solution for wireless channel estimation. Recent studies such as \cite{Jebur2021} and \cite{Hammed2023} have explored the use of ML in predicting channel characteristics with promising results, highlighting the superior performance of ML over conventional methods. In another study \cite{Maduranga2021}, the author demonstrated the effectiveness of Supervised ML (SML) for RSSI-based indoor localisation in IoT applications, underscoring the benefits of leveraging the benefits of ML over traditional estimation methods.
The researchers are also investigating the DL, advanced ML for channel estimation in IoT \cite{Thrane2020} \cite{Katagiri2019}. ANNs are the simplest techniques of DL-based methods. Implementing ANN-based models is considered accurate and resource-efficient in most applications, particularly suitable for LP-IoT environments. Although DL-based techniques have proved promising results in channel estimation in IoT, the high computational costs of these methods are a major concern. Therefore, these approaches require further improvement in channel estimation to increase accuracy and reduce computational resources.

Building on this foundation, this research focuses on developing and evaluating two specific ANN-based estimation methodologies designed for LP-IoT wireless channels: the Feature-based ANN model and the Sequence-based ANN model. The Feature-based approach leverages the extraction and learning of complex patterns inherent in raw RSSI data and estimates the future RSSI. On the other hand, the Sequence-based model selects a sequence of RSSI measurements using specific features to estimate upcoming channel conditions, hence improving the accuracy and efficiency of the estimation. By providing a comparative analysis with existing techniques, we underscore the potential of our ANN-based estimation approaches to improve the LP-IoT wireless communication systems substantially. The following are the main contributions to the field of wireless channel estimation LP-IoT devices:
\vspace{-1mm}
\begin{itemize}
    \item developed two distinct ANN-based models, namely the Feature-based and Sequence-based ANN models for wireless channel estimation in LP-IoT networks, signifying improvement in LP-IoT channel estimation.
    \item Collected empirical data utilising two LP-IoT devices, which were subsequently consolidated and engineered to assist with developing these models.
    \item Performed a comparative analysis of our suggested models with current research, traditional,  and other DL-based techniques, proving the superior performance of our models with the sophisticated percentage improvement in estimation error.
\end{itemize}

% \begin{figure}[h]
%     \begin{center}
%     \includegraphics[width=0.75\columnwidth]{storagelayers.jpg}
%     \caption{{The logical layered structure of blockchain implementation.}}
%     \label{fig:blockchainstoragecomponents}
%     \end{center}
% \end{figure}

\begin{figure}[h]
    \centering
    \includegraphics[width=0.6\linewidth]{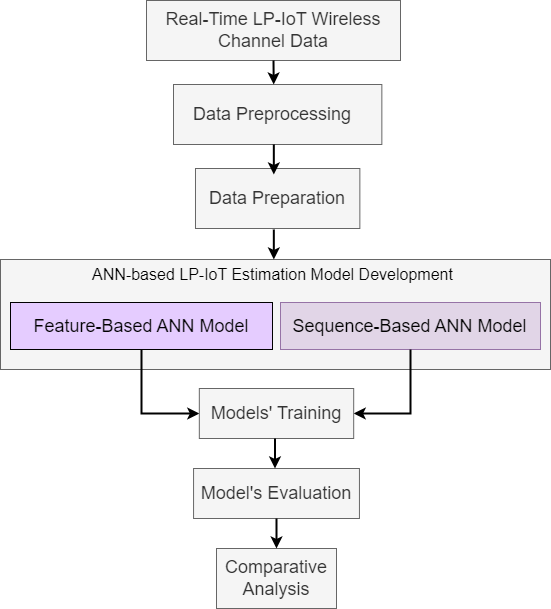}
    \caption{A block-based representation of the ANN-Based estimation process in LP-IoT systems.}
    \label{fig:SystemModel}
\end{figure}

\section{System Model and Problem Formulation}\label{sec:SysModel}
This section describes the system model employed in this article for LP-IoT wireless channel estimation and articulates the research problem formulation.

\subsection{System Model}
We begin developing the ANN-based channel estimation models for LP-IoT networks by collecting real-time channel data in an indoor environment. To collect the real data, we set up two LP-IoT devices, one acting as a transmitter and the other as a receiver in an indoor laboratory setting. The RSSI was then measured as the wireless channel metric in LP-IoT communication in environment modes for the model's development. The data collection was organised in three main scenarios (explained in Section \ref{Sec:Exp}): aggregated, prepossessed, and prepared for the model's development. Next, we developed two models based on ANN by analysing the RSSI as a wireless channel metric. These ANN models have been designed to estimate the RSSI in a dataset that presumably represents RSSI in an indoor IoT (Internet of Things) laboratory environment. The first model was developed utilising the environment features as input data for estimating the RSSI. Next, we developed the sequenced-based model in which we also utilised the aggregated data. However, we categorised it based on the environmental conditions, such as LoS and NLoS, varying receiver's location, and varying receiver's distance from the transmitter. Finally, we trained and evaluated both models and compared them with the traditional and DL-based estimation techniques. Table \ref{tab:Data} represents the randomly selected data from the main datasets to visualise the data arrangements.

\subsection{Problem Formulation}

In LP-IoT networks, the conventional wireless communication model is expressed by the equation $y=hx+n$. Here, $h$ represents the wireless channel between the LP-IoT transmitter and receiver, $x$ represents the transmitted signal, and $n$ denotes the noise in the system. The goal is to calculate the LP-IoT wireless channel gain in $dBm$, represented as RSSI and indicated as $y = |h|$. In the succeeding part, we will now move into a detailed development of our specifically developed models for the LP-IoT channel estimation.

\begin{figure*}[t]
    \centering
    \includegraphics[width=0.98\linewidth]{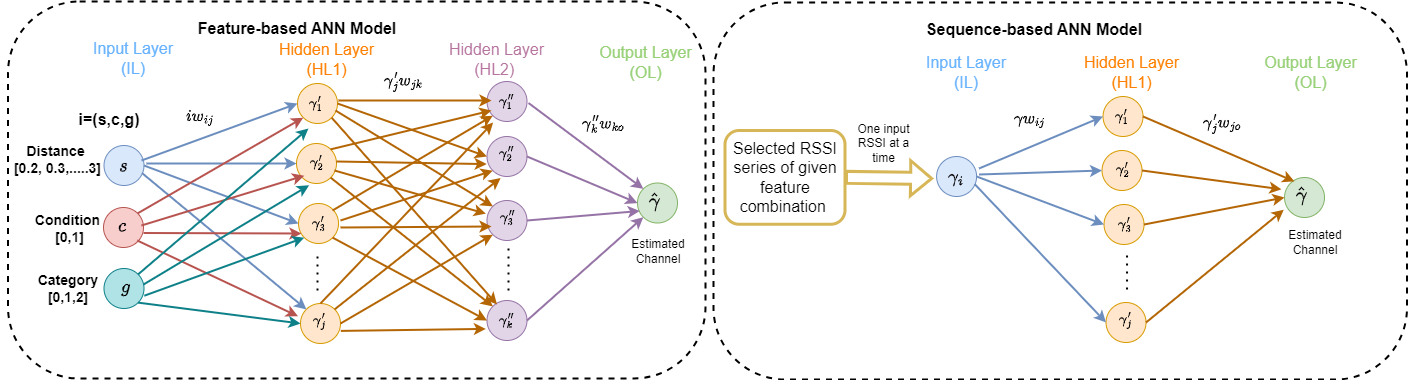}
    \caption{Visual representation of ANN-based LP-IoT estimation models.}
    \label{fig:Model-A}
\end{figure*}

\section{ANN Model Development}

\subsection{Feature-based ANN Model} \label{Sec:Featured-Mod}

\subsubsection{Phase I - Data Preparation} \label{SubSec:DataPrep}

Initially, we constructed the Feature-based model by aggregating the data from all three experimental scenarios and consolidating it into a unified spreadsheet. Subsequently, the environmental conditions derived from these experimental scenarios were incorporated into an ANN model as input features for future RSSI estimation. The input features consist of the distance, denoted as $s$, the Line-of-Sight (LoS) or Non-Line-of-Sight (NLoS) status, denoted as $c$, and the category determined by the location and distance between the transmitter and receiver, represented as $g$. The distance ranges from $0.1\vspace{1mm}m$ to $3\vspace{1mm}m$. The LoS is indicated by the number `0', while NLoS is marked by `1'. In the third feature, the fixed location is L1, categorised as `0'. The location varies from L2 to L12 and is categorised as `1'. The location and distance vary from L13 to L40 and are labelled as category `2' in the data spreadsheet. Table \ref{tab:Data} outlines data arrangement in the spreadsheet, which was employed as the input feature set for the Feature-based ANN model. In Table \ref{tab:Data}, the sequence column is disregarded at this stage because it has been employed in the Sequence-based model.

\subsubsection{Phase II - Model's Construction}

In the construction of our Feature-based ANN model, we designed it with four distinct layers: the Input Layer denoted as $IL$, two Hidden Layers labelled as $HL1$ and $HL2$, and the Output Layer as $OL$. The $IL$ receives the three primary input features: the distance $s$, the LoS and NLoS condition $c$, and the variable incorporating distance and location $g$. The $HL1$ and $HL2$ are equipped with 64 neurons each and play a pivotal role in eventually processing the inputs to yield the estimated RSSI at the $OL$. Throughout each layer, we applied the Rectified Linear Unit (ReLU) decision function to facilitate decision-making processes within the model. To assess the model's effectiveness, we undertook training sessions to observe its behaviour, followed by an evaluation using the remaining dataset. To accomplish this, we split the data into training and testing sets. The training set of $80\%$ and $20\%$ data was reserved for evaluating the model's performance on unseen data. Figure \ref{eq:Model-F} represents the configuration of the Feature-based ANN estimation model. A generalised mathematical representation of the constructed Feature-based model is described as follows:

%\vspace{-4mm}
\begin{equation} \label{eq:Model-F}
     \mathcal{M}_{(s,c,g)} \xrightarrow{\text{output}} \hat{\gamma},
\end{equation}
where $\mathcal{M}_{(s,c,g)}$ denotes the Feature-based ANN model, with input features distance as $s$, condition as $c$, and the category as $g$. The $\hat{\gamma}$ represents the estimated RSSI as output. The output $\hat{\gamma}$ in equation \ref{eq:Model-F} can be expanded as:  
%\vspace{-5mm}
\begin{equation} \label{eq:Model-FF}
    \hat{\gamma} \longleftarrow \delta_k(\gamma''_k) \longleftarrow \delta_j(\gamma'_j) \longleftarrow \gamma_i
\end{equation}

where $\gamma_{i}$ represents the input from $IL$ to $HL1$, whereas $ \gamma_{j}'$ and $\gamma_{k}''$ denotes the output at $HL1$, and $HL2$, respectively. The symbol $\hat{\gamma}$ represents the final estimated output at $OL$. The variables $i$, $j$, and $k$ indicate the number of neurons at $IL$, $HL1$, $HL2$ respectively. The $i$ contains three input features $s$,$c$, and $g$. The ReLU decision function integrated into the interconnected layers of the Feature-based ANN model is represented by $\delta_j$ and $\delta_k$.

\subsubsection{Phase III - Model Training}

To minimise errors in the constructed forward propagation model, backpropagation is required. For this, we used NAdam optimiser with a learning rate of $0.001$. We ran $1800$ epochs to reduce the estimation error. The Mean Squared Error (MSE) and Root Mean Squared Error (RMSE) were used as the loss function, also called the model's evaluation metrics. These metrics are essential for assessing the model's performance by calculating the discrepancies between the estimated and the actual values. Specifically, MSE is a straightforward statistic that computes the average squared difference between estimated and actual outcomes. The RMSE is obtained by calculating the square root of the MSE. The model's parameters were adjusted during the training phase to minimise the MSE loss function. The loss values were recorded for each epoch to monitor the training progress. After obtaining the minimum MSE during training, we calculated the RMSE. The mathematical expression of MSE and RMSE is depicted as follows: 
%\vspace{-2mm}
\begin{equation}
    MSE_\gamma=\frac{1}{N}(\gamma-\hat{\gamma})^2, \label{eq:MSE}
\end{equation}
where $N$ represents the total number of data samples, $\gamma$ denotes the collected RSSI, and $\hat{\gamma}$ denotes the estimated RSSI. The numeric training and testing results can be found in table \ref{tab:Comp-FeatureModel}. The RMSE can be expressed as:
%\vspace{-2mm}
\begin{equation}
    RMSE_\gamma=\sqrt{\frac{1}{N}(\gamma-\hat{\gamma})^2}. \label{eq:RMSE}
\end{equation}

\vspace{-3mm}
\begin{table}[t]
\small
\centering
\caption{A representation of the randomly selected data from the main aggregated data, representing the sequence used in Sequence-based ANN model.}
\label{tab:Data}
\begin{tabular}{|c|c|c|c|c|}
\hline
\rowcolor[HTML]{DAE8FC} 
\multicolumn{1}{|l|}{\cellcolor[HTML]{DAE8FC}\begin{tabular}[c]{@{}l@{}}RSSI \\ (dBm)\end{tabular}} &
\multicolumn{1}{l|}{\cellcolor[HTML]{DAE8FC}\begin{tabular}[c]{@{}l@{}}Distance \\ (0.2m - 3m)\end{tabular}} &
\multicolumn{1}{l|}{\cellcolor[HTML]{DAE8FC}\begin{tabular}[c]{@{}l@{}}Condition\\  (LoS/NLoS)\end{tabular}}&
\multicolumn{1}{l|}{\cellcolor[HTML]{DAE8FC}\begin{tabular}[c]{@{}l@{}}Location\\  (L1 to L40)\end{tabular}} &
\multicolumn{1}{l|}{\cellcolor[HTML]{DAE8FC}\begin{tabular}[c]{@{}l@{}}Sequence\end{tabular}}\\ \hline
-67.4 & 3   & LoS  & L1 & [3,0,0] \\ \hline
-65.2 & 3   & NLoS & L1 & [3,1,0] \\ \hline
-67   & 3   & LoS  & L2 & [3,0,1] \\ \hline
-57   & 3   & NLoS & L2 & [3,1,1] \\ \hline
-40   & 0.2 & LoS  & L13 & [0.2,0,2]\\ \hline
-47   & 0.2 & NLoS & L13 & [0.2,1,2] \\ \hline
-57   & 1.8 & NLoS & L29 & [1.8,1,2]\\ \hline
\end{tabular}
\end{table}

\subsection{Sequence-based ANN Model} \label{Sec:Seq-Mod}

\subsubsection{Phase I - Data Preparation}
This model is designed to estimate future RSSI values in a time series based on prior observations using the selected sequence of RSSI data. For this, we prepared the data to pick a sequence based on the distance, environment condition, varying location, and the distance between the transmitter and receiver (as explained in Section \ref{SubSec:DataPrep}). For the data preparation, we can see the data arrangement and the created sequence using the three columns `Distance', `Condition', and `Category' in the datasheet. We pick a series of RSSI values by selecting the arranged environment features such as [2,1,2]. Then, the selected sequence of a given length of RSSI values is trained and evaluated to estimate the next RSSI values in the time series manner. This unique model is extendable to new features based on distance, condition, and category by making suitable adjustments in the environment features.

\subsubsection{Phase II - Model's Construction}
After performing the data preparation method, we configured the Sequence-based ANN model in three layers: the Input Layer (IL), the Hidden Layer (HL1), and the Output Layer (OL). From the series of selected RSSI data, the IL processes one data point at a time. The HL1 comprises 64 neurons, while the OL generates a single estimation at each instance. This structured approach allows the estimation of the LP-IoT channel data under different environmental conditions individually. This model does not utilise the entire dataset; instead, it picks the sequence, rendering it a highly efficient model. The generalised equation of the proposed approach can be mathematically expressed as:

\vspace{-3mm}
\begin{equation} \label{eq:Seq-Model}
    \mathcal{S}_{(s,c,g)}^\gamma \xrightarrow{\text{selected sequence}} \mathcal{M}_{sq} \xrightarrow{\text{output}} \hat{\gamma},
\end{equation}

where $\mathcal{S}_{(s,c,g)}^\gamma$ represents the RSSI sequence selected by the different combinations of input features. The variables s,c, and g denote the data's specific features, such as distance, condition, and category. The $\gamma$ indicates the corresponding measured RSSI value the model aims to predict. $\mathcal{M}_{sq}$ denotes the Sequence-based ANN model that processes the selected sequence picked up by the input features, and $\hat{\gamma}$ represents the estimated RSSI, the model's output. The model $\mathcal{M}_{sq}$ can be expressed in its expanded version as:

\vspace{-3mm}
\begin{equation} \label{eq:Model-SS}
    \hat{\gamma} \longleftarrow \delta_j(\gamma'_j) \longleftarrow \gamma_i,
\end{equation}

where $\gamma_i$ represents the input data for processing, $\delta_j(\gamma'_j)$ denotes the transformation operation at $HL1$ with the ReLU decision function, denoted as $\delta_j$, and the $\hat{\gamma}$ represents the estimated output at the $OL$.

\subsubsection{Phase III - Model Training}
We trained a Sequence-based ANN model, emphasising a lightweight and straightforward approach to selecting and processing the features. We adopted the ADAM optimiser with a learning rate of 0.01 for our training process to facilitate this. In the Sequence-based ANN model, the choice of ADAM over NADAM was deliberate as we aimed to maintain the model's simplicity and avoid unnecessary complexity. The ADAM has lower computational demands than the NADAM. Furthermore, we implemented a dropout rate of $0.5$ to mitigate over-fitting while preserving the model's ability to generalise from the training data. The training was extended until 200 epochs to achieve the minimum error rate. Model's performance is evaluated using MSE and RMSE equations \ref{eq:MSE} and \ref{eq:RMSE} respectively.

\section{Experimental Data Collection} \label{Sec:Exp}

The experimental framework was established in an indoor laboratory at Charles Sturt University. The experimental test bed was designed to simulate real-world IoT communications using two Waspmote devices for transmitting and receiving signals. The device setup and configuration can be found in our earlier research \cite{Samrah2024}. We measured RSSI as the primary channel metric and analysed it to gauge wireless channel performance. We designed three scenarios to depict the typical IoT wireless channel: LoS and NLoS environment conditions. 

The first designed scenario examined the stationary devices at fixed distances to establish a baseline for signal strength in unobstructed and obstructed arrangements, using common indoor items as physical barriers. Initially, the LoS setup was arranged on a centre table with the transmitter and receiver at a fixed distance of $3$ metres apart. This setting was carried out to ensure a direct LoS and evaluate the basic communication performance in a controlled environment. Consequently, the physical barriers were introduced to create an NLoS condition between the transmitter and receiver. In the NLoS setup, the devices were placed on opposite sides of a standard wooden lecture table and equipped with power points and a screen. This table was the obstructing medium, including an additional whiteboard and a chair.

The second scenario examined the impact of varying the receiver's position within the laboratory to ascertain the influence of location variability on signal transmission. To examine the LoS condition, We methodically positioned the receiver at different locations on a specified table, all within a constant distance of $3\hspace{1mm}m$. To simulate the NLoS condition, we incorporated common laboratory objects such as a small CPU, a book, and a hand sanitiser, thus producing an authentic NLoS environment. The receiver's position was slightly adjusted across $11$ distinct places within the setup table to allow the influence of spatial dynamics on signal transmission. The third scenario expanded this study by examining varying distances between the transmitter and receiver, considering both LoS and NLoS conditions. We set up the LoS condition by measuring RSSI at an initial distance of $0.2\hspace{1mm}m$ between the transmitter and receiver. We then systematically increased this distance by $0.1\hspace{1mm}m$ increments, up to a maximum distance of $2.9\hspace{1mm}m$. 

In the initial setup, the experiment was carried out over a prolonged duration of more than three hours, allowing for the acquisition of 10,000 samples for each condition. In scenarios 2 and 3, each specific location and distance were tested for a short experimental period of around five minutes. During this period, the number of samples fluctuated between $220$ and $260$, mostly due to occurrences of random packet loss. The total number of RSSI samples obtained by aggregating data from all scenarios is $43163$.While preparing the dataset for model implementation, we thoroughly examined the data and performed the cleansing to identify and eliminate any empty entries.

\begin{figure*}[t]
    \centering
    \includegraphics[width=0.88\linewidth]{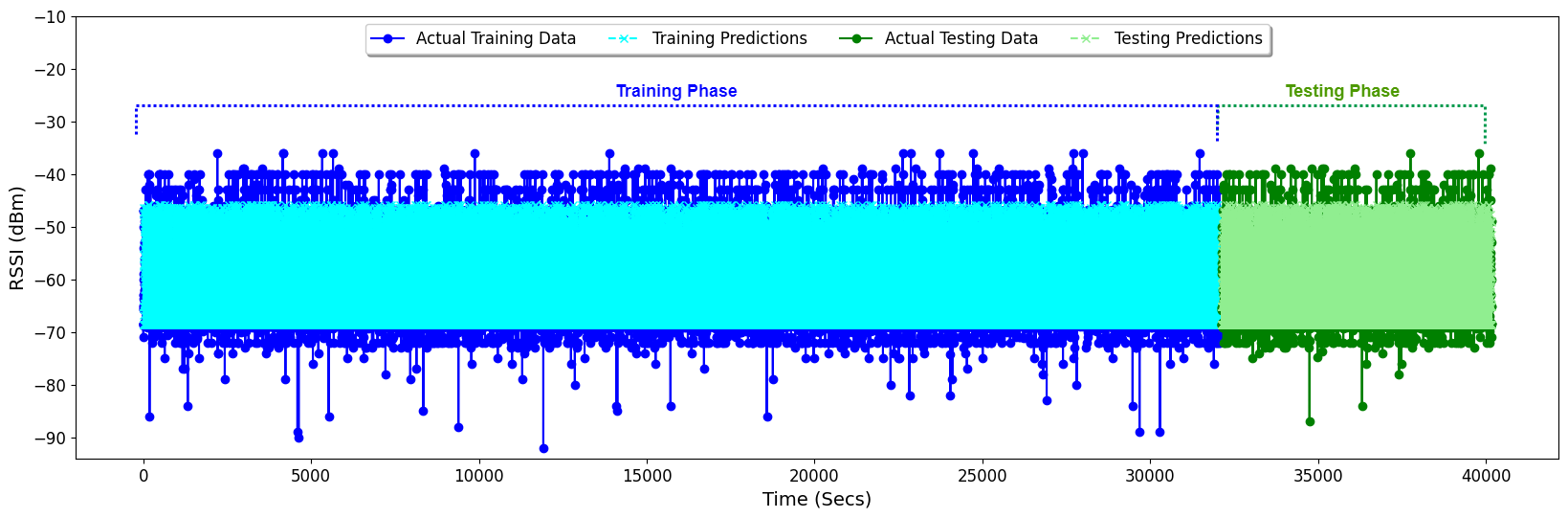}
    \caption{Visual representation of the results of Feature-based ANN model for LP-IoT wireless channel estimation.}
    \label{fig:Feature-Model}
\end{figure*}

\begin{figure*}[!h]
    \centering
    % First row
    \begin{subfigure}{0.49\linewidth}
        \includegraphics[width=\linewidth]{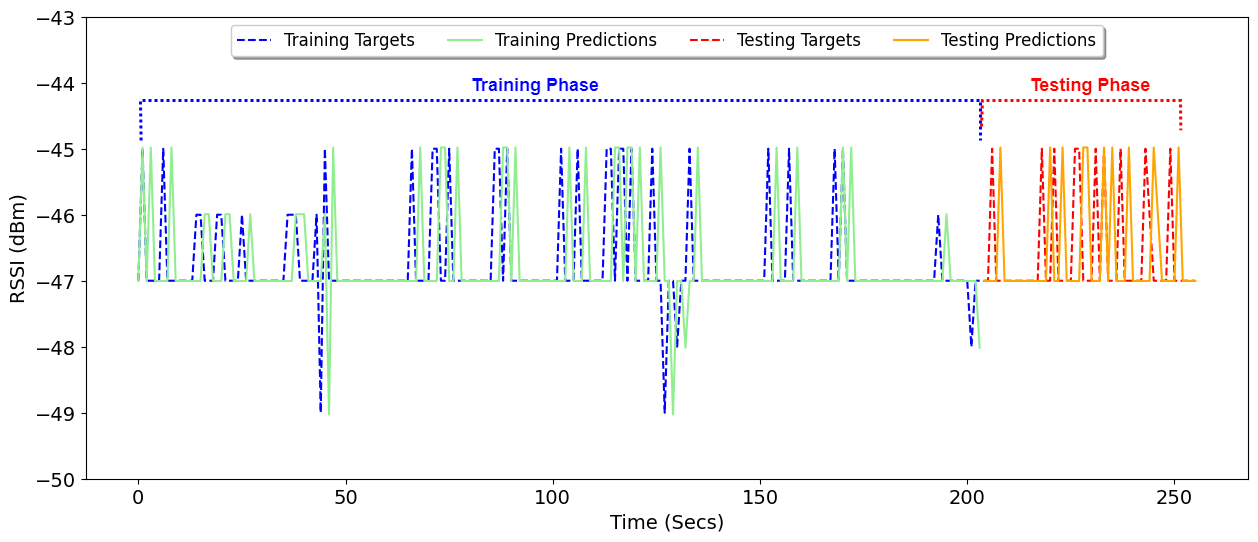}
        \caption{The results of selected Sequence using the features [0.5, 0, 2].}
        \label{Fig:S3-C1-0.5mResult-a}
    \end{subfigure}
    \hfill
    \begin{subfigure}{0.49\linewidth}
        \includegraphics[width=\linewidth]{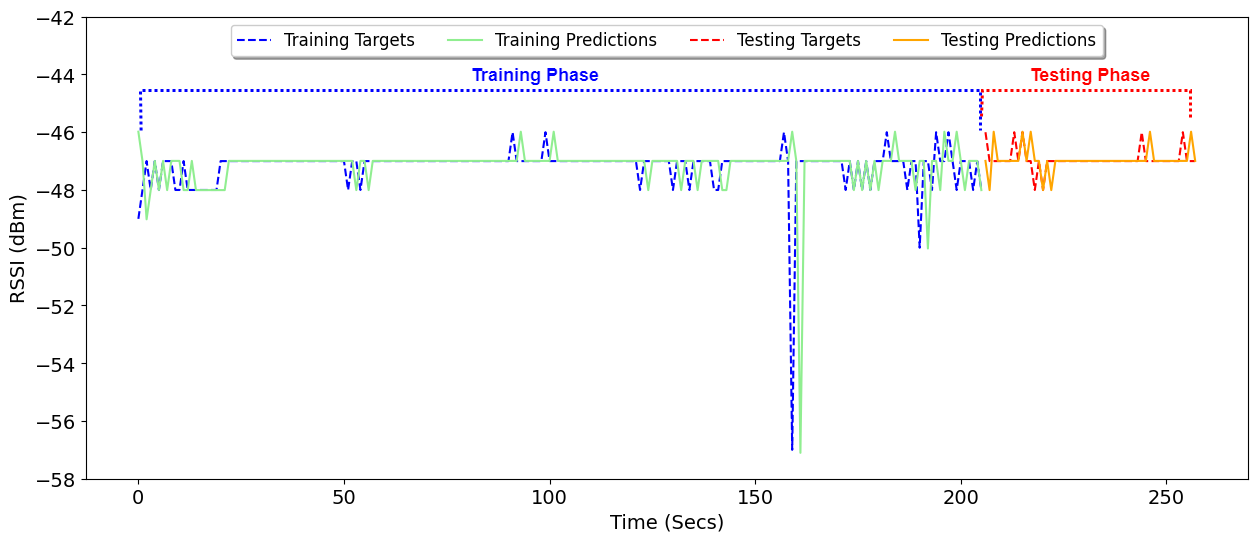}
        \caption{The results of selected sequence using the features [0.5, 1, 2].}
        \label{Fig:S3-C2-0.5mResult-b}
    \end{subfigure}
    
    % Second row
    \begin{subfigure}{0.49\linewidth}
        \includegraphics[width=\linewidth]{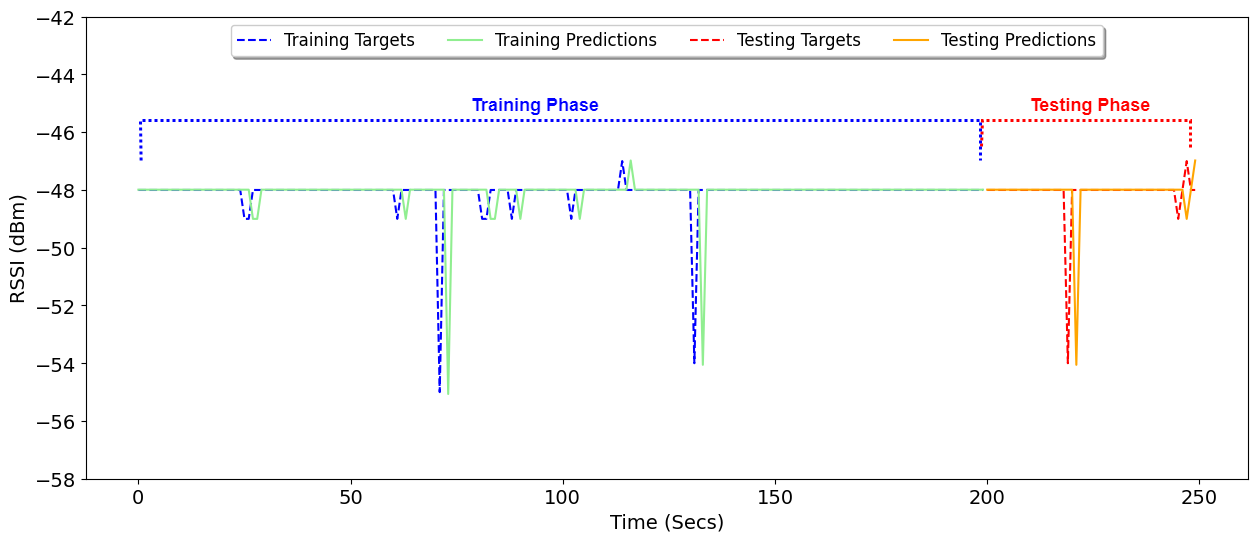}
        \caption{The results of selected sequence using the features [1, 0, 2].}
        \label{Fig:S3-C1-1mResult-c}
    \end{subfigure}
    \hfill
    \begin{subfigure}{0.49\linewidth}
       \includegraphics[width=\linewidth]{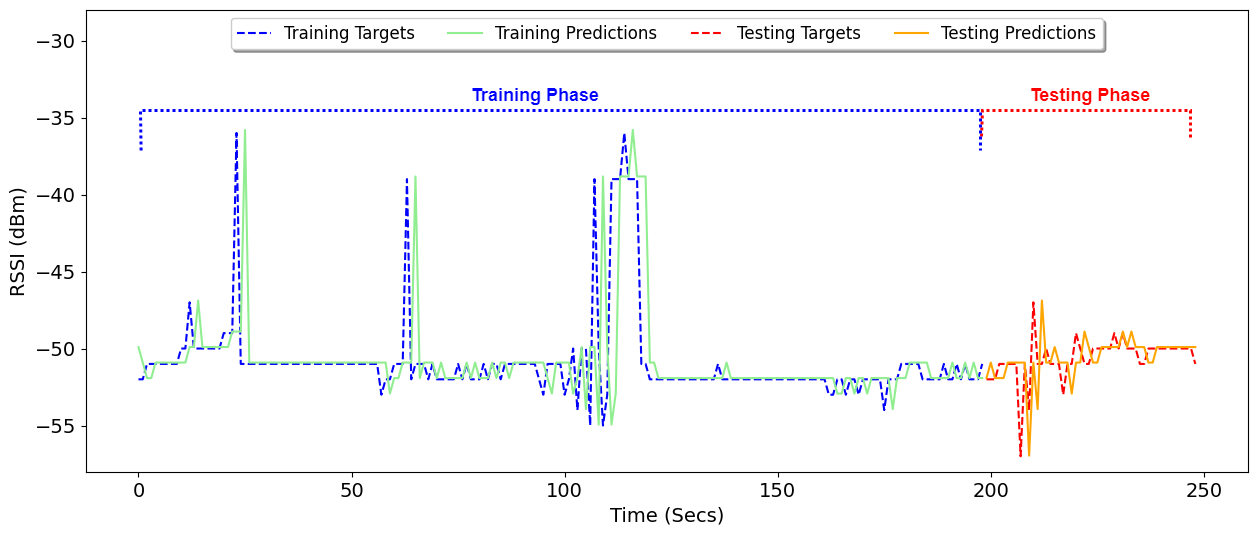}
        \caption{The results of selected sequence using the features [1, 1, 2].}
        \label{Fig:S3-C2-1mResult-d}
    \end{subfigure}
    
    % Third row
    \begin{subfigure}{0.49\linewidth}
        \includegraphics[width=\linewidth]{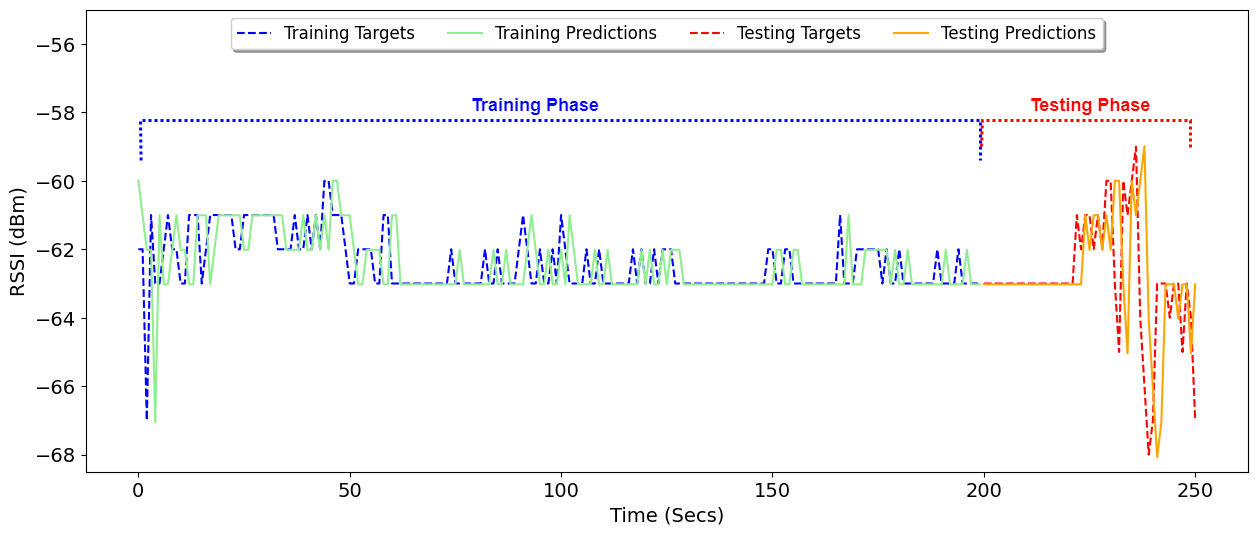}
        \caption{The results of selected sequence using the features [2, 0, 2].}
        \label{Fig:S3-C1-2mResult-e}
    \end{subfigure}
    \hfill
    \begin{subfigure}{0.49\linewidth}
        \includegraphics[width=\linewidth]{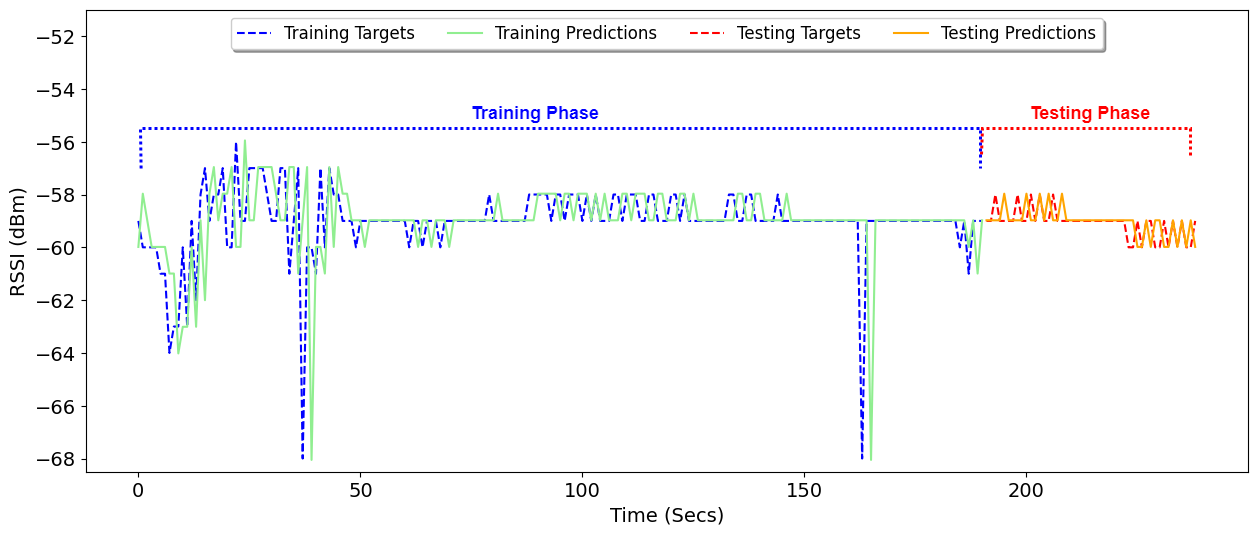}
        \caption{The results of selected sequence using the features [2, 1, 2].}
        \label{Fig:S3-C2-2mResult-f}
    \end{subfigure}
    \caption{Visual representation of the results of Sequence-based ANN model for LP-IoT wireless channel estimation.}
    \label{fig:comprehensive-results}
\end{figure*}

\section{Performance Evaluation and Discussion} \label{sec:Results}
In this research, we implemented two models: A Feature-based ANN model and a Sequence-based ANN model. The Feature-based ANN model integrates the LP-IoT environment features and estimates the RSSI based on these features. The evaluation outcome of this model demonstrates a substantial performance improvement when compared to the traditional Regression model and the other DL-based techniques, RNN and LSTM, throughout both the training and testing phases.  

In the training phase, the Feature-based ANN model attains an MSE of $5.91\hspace{1mm}dBm$ and an RMSE of $2.43\hspace{1mm}dBm$. Upon evaluation, the model exhibits a slight reduction in error with MSE of $5.30\hspace{1mm}dBm$ and RMSE of $2.30\hspace{1mm}dBm$. In contrast, the traditional Regression model shows elevated error rates in both training and testing, with an MSE of $8.62\hspace{1mm}dBm$ and RMSE of $2.93\hspace{1mm}dBm$. These results indicate that the Feature-based ANN model possesses a robust generalisation ability as compared to the traditional estimation approach. Subsequently, we evaluate this model in comparison to other DL-based models, such as RNN and LSTM. The models also possess higher MSE and RMSE in the training and testing phases than the Feature-based ANN model. Table \ref{tab:Comp-FeatureModel} presents a comparative analysis that visually demonstrates the superior performance of the Feature-based ANN model in comparison to traditional and other DL-based approaches.

\begin{table}[t]
\caption{The comparison of Feature-based ANN model with traditional and other DL-based techniques.}
\label{tab:Comp-FeatureModel}
\begin{tabular}{|l|ll|ll|}
\hline
\multicolumn{1}{|c|}{\multirow{2}{*}{\textbf{Estimation Models}}} & \multicolumn{2}{l|}{\textbf{Training}}            & \multicolumn{2}{l|}{\textbf{Testing}}             \\ \cline{2-5} 
\multicolumn{1}{|c|}{}                                            & \multicolumn{1}{l|}{\textbf{MSE}} & \textbf{RMSE} & \multicolumn{1}{l|}{\textbf{MSE}} & \textbf{RMSE} \\ \hline
Feature-based ANN Model & \multicolumn{1}{l|}{5.91} & 2.43 & \multicolumn{1}{l|}{5.30} & 2.30 \\ \hline
Regression Model        & \multicolumn{1}{l|}{8.62} & 2.93 & \multicolumn{1}{l|}{8.62} & 2.93 \\ \hline
RNN                     & \multicolumn{1}{l|}{7.48} & 2.72 & \multicolumn{1}{l|}{6.72} & 2.59 \\ \hline
LSTM                    & \multicolumn{1}{l|}{51.92} & 7.20 & \multicolumn{1}{l|}{51.44} & 7.17 \\ \hline
\end{tabular}
\end{table}

\begin{table}[t]
\caption{Comparison of proposed Sequence-based ANN estimation model with other DL-based techniques for LP-IoT channel estimation.}
\label{tab:Sequence-based-Comparison}
\begin{tabular}{|lllllll|}
\hline
\multicolumn{1}{|l|}{} &
  \multicolumn{2}{c|}{\textbf{Training}} &
  \multicolumn{2}{c|}{\textbf{Testing}} &
  \multicolumn{1}{c|}{} &
  \multicolumn{1}{c|}{} \\ \cline{2-5}
\multicolumn{1}{|l|}{\multirow{-2}{*}{\textbf{\begin{tabular}[c]{@{}l@{}}Selected \\ Sequence\end{tabular}}}} &
  \multicolumn{1}{l|}{\textbf{MSE}} &
  \multicolumn{1}{l|}{\textbf{RMSE}} &
  \multicolumn{1}{l|}{\textbf{MSE}} &
  \multicolumn{1}{l|}{\textbf{RMSE}} &
  \multicolumn{1}{c|}{\multirow{-2}{*}{\textbf{\begin{tabular}[c]{@{}c@{}}Training\\ Time(s)\end{tabular}}}} &
  \multicolumn{1}{c|}{\multirow{-2}{*}{\textbf{\begin{tabular}[c]{@{}c@{}}Testing \\ Time(s)\end{tabular}}}} \\ \hline
\multicolumn{7}{|c|}{\cellcolor[HTML]{DAE8FC}\textbf{Sequence-based ANN Estimation Model}} \\
\hline
\multicolumn{1}{|l|}{\textbf{{[}3, 0, 0{]}}} &
  \multicolumn{1}{l|}{0.27} &
  \multicolumn{1}{l|}{0.52} &
  \multicolumn{1}{l|}{0.19} &
  \multicolumn{1}{l|}{0.43} &
  \multicolumn{1}{l|}{1.022} &
  0.0015 \\ \hline
\multicolumn{1}{|l|}{\textbf{{[}3, 1, 0{]}}} &
  \multicolumn{1}{l|}{0.39} &
  \multicolumn{1}{l|}{0.62} &
  \multicolumn{1}{l|}{0.38} &
  \multicolumn{1}{l|}{0.62} &
  \multicolumn{1}{l|}{0.902} &
  0.0013 \\ \hline
\multicolumn{1}{|l|}{\textbf{{[}0.5, 0, 2{]}}} &
  \multicolumn{1}{l|}{0.90} &
  \multicolumn{1}{l|}{0.95} &
  \multicolumn{1}{l|}{1.43} &
  \multicolumn{1}{l|}{1.19} &
  \multicolumn{1}{l|}{0.739} &
  0.0005 \\ \hline
\multicolumn{1}{|l|}{\textbf{{[}0.5, 1, 2{]}}} &
  \multicolumn{1}{l|}{1.45} &
  \multicolumn{1}{l|}{1.20} &
  \multicolumn{1}{l|}{0.21} &
  \multicolumn{1}{l|}{0.46} &
  \multicolumn{1}{l|}{0.381} &
  0.0004 \\ \hline
\multicolumn{1}{|l|}{\textbf{{[}1, 0, 2{]}}} &
  \multicolumn{1}{l|}{0.94} &
  \multicolumn{1}{l|}{0.96} &
  \multicolumn{1}{l|}{1.57} &
  \multicolumn{1}{l|}{1.25} &
  \multicolumn{1}{l|}{0.422} &
  0.0005 \\ \hline
\multicolumn{1}{|l|}{\textbf{{[}2, 0, 2{]}}} &
  \multicolumn{1}{l|}{0.91} &
  \multicolumn{1}{l|}{0.95} &
  \multicolumn{1}{l|}{4.36} &
  \multicolumn{1}{l|}{2.08} &
  \multicolumn{1}{l|}{0.369} &
  0.0009 \\ \hline
\multicolumn{1}{|l|}{\textbf{{[}2, 1, 2{]}}} &
  \multicolumn{1}{l|}{2.86} &
  \multicolumn{1}{l|}{1.69} &
  \multicolumn{1}{l|}{0.35} &
  \multicolumn{1}{l|}{0.59} &
  \multicolumn{1}{l|}{0.360} &
  0.0010 \\ \hline
\multicolumn{7}{|c|}{\cellcolor[HTML]{DAE8FC}\textbf{Recurrent Neural Networks (RNNs)-based Estimation Model}} \\ \hline
\multicolumn{1}{|l|}{\textbf{{[}3, 0, 0{]}}} &
  \multicolumn{1}{l|}{2.89} &
  \multicolumn{1}{l|}{1.70} &
  \multicolumn{1}{l|}{2.50} &
  \multicolumn{1}{l|}{1.58} &
  \multicolumn{1}{l|}{55.979} &
  0.0013 \\ \hline
\multicolumn{1}{|l|}{\textbf{{[}3, 1, 0{]}}} &
  \multicolumn{1}{l|}{1.32} &
  \multicolumn{1}{l|}{1.14} &
  \multicolumn{1}{l|}{1.79} &
  \multicolumn{1}{l|}{1.33} &
  \multicolumn{1}{l|}{59.129} &
  0.0016 \\ \hline
\multicolumn{1}{|l|}{\textbf{{[}0.5, 0, 2{]}}} &
  \multicolumn{1}{l|}{0.50} &
  \multicolumn{1}{l|}{0.70} &
  \multicolumn{1}{l|}{0.64} &
  \multicolumn{1}{l|}{0.80} &
  \multicolumn{1}{l|}{24.000} &
   0.0052\\ \hline
\multicolumn{1}{|l|}{\textbf{{[}0.5, 1, 2{]}}} &
  \multicolumn{1}{l|}{0.68} &
  \multicolumn{1}{l|}{0.82} &
  \multicolumn{1}{l|}{0.18} &
  \multicolumn{1}{l|}{0.43} &
  \multicolumn{1}{l|}{28.050} &
  0.0011 \\ \hline
\multicolumn{1}{|l|}{\textbf{{[}1, 0, 2{]}}} &
  \multicolumn{1}{l|}{0.45} &
  \multicolumn{1}{l|}{0.67} &
  \multicolumn{1}{l|}{0.73} &
  \multicolumn{1}{l|}{0.85} &
  \multicolumn{1}{l|}{0.2474} &
  0.0003 \\ \hline
\multicolumn{1}{|l|}{\textbf{{[}2, 0, 2{]}}} &
  \multicolumn{1}{l|}{1.16} &
  \multicolumn{1}{l|}{1.07} &
  \multicolumn{1}{l|}{4.23} &
  \multicolumn{1}{l|}{2.05} &
  \multicolumn{1}{l|}{0.2506} &
  0.0008 \\ \hline
\multicolumn{1}{|l|}{\textbf{{[}2, 1, 2{]}}} &
  \multicolumn{1}{l|}{2.03} &
  \multicolumn{1}{l|}{1.42} &
  \multicolumn{1}{l|}{0.45} &
  \multicolumn{1}{l|}{0.67} &
  \multicolumn{1}{l|}{0.2701} &
  0.0010 \\ \hline
\multicolumn{7}{|c|}{\cellcolor[HTML]{DAE8FC}\textbf{Long Short-Term Memory (LSTM)-based Estimation Model}} \\ \hline
\multicolumn{1}{|l|}{\textbf{{[}3, 0, 0{]}}} &
  \multicolumn{1}{l|}{21.93} &
  \multicolumn{1}{l|}{4.68} &
  \multicolumn{1}{l|}{21.30} &
  \multicolumn{1}{l|}{4.61} &
  \multicolumn{1}{l|}{151.55} &
  0.4225 \\ \hline
\multicolumn{1}{|l|}{\textbf{{[}3, 1, 0{]}}} &
  \multicolumn{1}{l|}{35.39} &
  \multicolumn{1}{l|}{5.94} &
  \multicolumn{1}{l|}{37.93} &
  \multicolumn{1}{l|}{6.15} &
  \multicolumn{1}{l|}{166.70} &
  0.4226 \\ \hline
\multicolumn{1}{|l|}{\textbf{{[}0.5, 0, 2{]}}} &
  \multicolumn{1}{l|}{0.50} &
  \multicolumn{1}{l|}{0.70} &
  \multicolumn{1}{l|}{0.64} &
  \multicolumn{1}{l|}{0.80} &
  \multicolumn{1}{l|}{31.840} &
   0.1100\\ \hline
\multicolumn{1}{|l|}{\textbf{{[}0.5, 1, 2{]}}} &
  \multicolumn{1}{l|}{0.68} &
  \multicolumn{1}{l|}{0.82} &
  \multicolumn{1}{l|}{0.18} &
  \multicolumn{1}{l|}{0.43} &
  \multicolumn{1}{l|}{33.560} &
  0.1200 \\ \hline
\multicolumn{1}{|l|}{\textbf{{[}1, 0, 2{]}}} &
  \multicolumn{1}{l|}{0.46} &
  \multicolumn{1}{l|}{0.68} &
  \multicolumn{1}{l|}{0.74} &
  \multicolumn{1}{l|}{0.86} &
  \multicolumn{1}{l|}{5.77} &
  0.4385 \\ \hline
\multicolumn{1}{|l|}{\textbf{{[}2, 0, 2{]}}} &
  \multicolumn{1}{l|}{17.78} &
  \multicolumn{1}{l|}{4.21} &
  \multicolumn{1}{l|}{24} &
  \multicolumn{1}{l|}{4.89} &
  \multicolumn{1}{l|}{0.835} &
  0.4386 \\ \hline
\multicolumn{1}{|l|}{\textbf{{[}2, 1, 2{]}}} &
  \multicolumn{1}{l|}{7.83} &
  \multicolumn{1}{l|}{2.79} &
  \multicolumn{1}{l|}{6.54} &
  \multicolumn{1}{l|}{2.55} &
  \multicolumn{1}{l|}{0.863} &
  0.4385 \\ \hline
\end{tabular}
\end{table}

Next, we acquired the results of the Sequence-based ANN model using sequences of different feature combinations, such as [3,0,0], [3,1,0], [0.5,0,2], [0.5,1,2], [1,0,2], [2,0,2] and [2,1,2]. From Table \ref{tab:Sequence-based-Comparison}, we can visualise the notable efficiency of our Sequence-based ANN model in comparison to the RNN and LSTM models using different selected sequences. Specifically, using the sequence with features [3,0,0], the Sequence-based ANN model showed promising results with an MSE of $0.27\hspace{1mm}dBm$ and RMSE of $0.52\hspace{1mm}dBm$ during its training phase. It further reduced the testing error with an MSE of $0.19\hspace{1mm}dBm$ and RMSE of $0.43\hspace{1mm}dBm$. Additionally, the training and testing time of $1.022\hspace{1mm}$ and $0.0015\hspace{1mm}$ seconds represents the minimal computational overhead. Conversely, the RNN-based model achieves an MSE of $2.89\hspace{1mm}dBm$ and RMSE of $1.70\hspace{1mm}dBm$ during the training phase. However, there is a slight improvement in the testing phase, with MSE $2.50\hspace{1mm}dBm$ and RMSE $1.58\hspace{1mm}dBm$. Similarly, the LSTM-based model exhibits a stark decline in performance, with an MSE of $21.93\hspace{1mm}dBm$ in training and $21.30\hspace{1mm}dBm$ in testing. Also, we noticed that the RNN and LSTM models have a significantly higher computational cost during the training phase, with $55.979\hspace{1mm}$seconds and $151.55\hspace{1mm}$seconds, which reflect the inefficiencies in these models. Through this evaluation, we exemplify the robustness of the Sequence-based ANN model with a minimal training period, in contrast to the traditional and other DL-based techniques.

Lastly, we compare our developed models with the existing research. The findings demonstrate that our suggested approaches attain remarkable precision in channel estimation, with the improvement of $88\%$ of Feature-based model with an MSE of $5.30\hspace{1mm}dBm$, and $97\%$ of Sequence-based model with an average MSE of six sequences of $1.15\hspace{1mm}dBm$ over existing research in \cite{Raj2021} with an average MSE of six datasets of $45.25\hspace{1mm}dBm$. This comprehensive analysis shows how our models outperform the traditional, DL-based, and other existing techniques. However, it also underlines the importance of continued research in exploring hybrid models. The hybrid models, that combine several strategies may enhance the estimation efficiency, particularly in handling the unpredictability and complexity of the LP-IoT wireless channel.

\section{Conclusion and Future Work}
This research constructed and evaluated two ANN-based estimation models for LP-IoT wireless systems: the Feature-based ANN and Sequence-based ANN models. The Feature-based model utilises three distinct features of the channel to estimate RSSI. The Sequence-based model was designed to estimate future RSSI values using a sequence of past RSSI measurements picked from the combination of three features. From the table \ref{tab:Comp-FeatureModel}, we illustrated the Feature-based model's remarkable accuracy and efficiency compared to traditional and other DL-based estimation methods such as RNN and LSTM. Similarly, the results depicted in table \ref{tab:Sequence-based-Comparison} affirm the potential of the Sequence-based ANN model in enhancing LP-IoT channel estimation, offering a significant improvement over DL-based approaches in terms of simplicity and efficiency. For future research, the real-time adaptability of our models to diverse environmental conditions, such as multi-room and multi-floor environments, along with improved scalability and computational efficiency, are the key areas for investigation.

 \nocite{*} 
\bibliographystyle{unsrt}
%\bibliographystyle{IEEEtran}
%\bibliography{ICECET.bbl}

\bibliography{RefList.bib}  %%% 

%Uncomment this line and comment out the ``thebibliography'' section below to use the external .bib file (using bibtex) .





% \end{thebibliography}

%\balance
%\bibliographystyle{IEEEtran}
%{\footnotesize \bibliography{RefList.bib}}

%\vskip 0pt plus -1fil

\end{document}